\documentclass{aa} 
\usepackage{graphicx}
\usepackage{txfonts}
\usepackage{longtable}
\usepackage{hyperref}

\usepackage{rotating}
\usepackage{booktabs}

\input epsf.sty

\begin{document}

\title{The properties of active galaxies at the extreme of Eigenvector 1}

\author{M. \' Sniegowska \inst{1,2}
        \and
      B. Czerny\inst{1}
      \and
      B. You\inst{4,3}
      \and
      S. Panda\inst{1,3}
      \and
      J.-M. Wang\inst{5}
      \and
      K. Hryniewicz\inst{3}
      \and
      C. Wildy\inst{1}
        }

 \offprints{M. \' Sniegowska marzena.sniegowska@student.uw.edu.pl}

\institute{Center for Theoretical Physics, Polish Academy of Sciences, Al. Lotnik\' ow 32/46, 02-668 Warsaw, Poland\\
     \and{Warsaw University Observatory, Al. Ujazdowskie 4, 00-478 Warsaw, Poland}\\
     \and{Copernicus Astronomical Center, Bartycka 18, 00-716 Warsaw, Poland} \\
     \and{School of Physics and Technology, Wuhan University, Wuhan 430072, China} \\
     \and{Key Laboratory for Particle Astrophysics, Institute of High Energy Physics, Chinese Academy of Sciences, 19B Yuquan Road, Beijing 100049, China}
     }

\abstract
{Eigenvector 1 (EV1) is the formal parameter which allows the introduction of some order in the properties of the unobscured type 1 active galaxies.}
{We aim to understand the nature of this parameter by analyzing the most extreme examples of quasars with the highest possible values of the corresponding eigenvalues $R_{Fe}$.}
{We selected the appropriate sources from the Sloan Digital Sky Survey (SDSS) and performed detailed modeling, including various templates for the Fe II pseudo-continuum and the starlight contribution to the spectrum.}  
{Out of 27 sources with $R_{Fe}$ larger than 1.3 and with the measurement errors smaller than 20\%  selected from the SDSS quasar catalog, only six sources were confirmed to have a high value of $R_{Fe}$, defined as being above 1.3. All other sources have $an R_{Fe}$ of approximately 1. Three of the high $R_{Fe}$ objects have a very narrow H$\beta$ line, below 2100 km s$^{-1}$ but three sources have broad lines, above 4500 km s$^{-1}$, that do not seem to form a uniform group, differing considerably in black hole mass and Eddington ratio; they simply have a very similar EW([OIII]5007) line. Therefore, the interpretation of the EV1 remains an open issue.}
{}

\authorrunning{\' Sniegowska et al.}
\titlerunning{Extreme active galactic nuclei}
\maketitle

\section{Introduction}

Active galactic nuclei (AGN) show a broad range of properties when their emission lines and wide-range spectra distributions are studied. Here we concentrate on the radio-quiet type 1 sources, however they do not appear to form a uniform sample, as expected in the simplest version of the AGN unification model (e.g., Peterson 1989). Theoretically, we still expect some dependence on the viewing angle, although in type 1 objects we have a clear view of the nucleus and therefore a reliable constraint on the black hole mass, Eddington ratio, and the black hole spin. Observationally, principal component analysis (PCA) showed that there is a hidden single parameter which corresponds to a significant part of the dispersion in the measured parameters. This idea, first pointed out by Boroson \& Green (1992), has been subsequently developed by numerous authors (e.g., Sulentic et al. 2000, Marziani et al. 2001; Kuraszkiewicz et al. 2009; Shen \& Ho 2014).  The underlying Eigenvector 1 (EV1) was a linear combination of several parameters, being mostly driven by the anticorrelation between the strengths of optical Fe~{\sc ii} and [O~{\sc iii}] $\lambda$5007.

However, the relation between the EV1 and the theoretically motivated parameters is not yet clear.
The Eddington ratio is most frequently favored (e.g., Sulentic et al. 2000), as already proposed by Boroson \& Green in 
the original paper on that subject. Shen \& Ho (2014) argued convincingly that the viewing angle is not the main driver of EV1, 
and that it simply provides a dispersion to the quasar main sequence. However, they did not uniquely solve the problem of the physical parameter
varying along the quasar main sequence, as they showed that the Eddington ratio increases with EV1 while the black hole mass decreases. This issue is important for better understanding of the AGN, as well as for possible cosmological applications 
(Marziani \& Sulentic 2014).

We thus selected sources with extreme properties populating the Shen \& Ho (2014) diagram. 
We selected sources with the measurement of Fe II optical flux and broad H$\beta$ flux with accuracy better than 20\%, 
and reanalyzed them much more carefully with the aim of confirming their extreme properties.

\section{Method}

\subsection{Observational data}

We select the extreme EV1 active galaxies from the Shen et al. (2011) catalog, taking into consideration only those objects with the Fe II optical flux to 
H$\beta$ ratio, $R_{Fe}$, measured with the errors below 20\%. We set the minimum value of the ratio $R_{Fe}$ at 1.3, in order to get a moderately large sample of 27 objects (selection at $R_{Fe} > 2.0$ returned only 9 objects).
The average value of $R_{Fe}$ in the Shen et al. (2011) catalog is 0.64, if only objects with high  quality data are included.
The list is given in Table~\ref{tab:objects}. The spectra were corrected for Galactic reddening using the Cardelli et al. (1989) 
extinction curve and taking the required amount of reddening from NASA/IPAC Extragalactic Database (NED).
\begin{table*}
\caption{Extreme EV1 objects selected for the study. Parameters taken from Shen et al. (2011) catalog.}  
\label{tab:objects}      
\centering                          
\begin{tabular}{l r r r  r  r  r  r  r  r }        
\hline\hline      
Name    &  RA  &  Dec  & FWHM(H$\beta$)&   EW(FeII) & EW(H$\beta$) & $R_{Fe}$ & redshift &  $\log L_{5100}$   & $\log M_{BH}$   \\
         &  &    &  km s$^{-1}$ & & \AA & \AA    \\
\hline
014723.27+144320.9 & 26.847  & 14.722 & 1349 & 37.2 & 59.3 & 1.59 & 0.4327   & 44.690  & 7.51     \\
020028.37-093859.0 & 30.118  & -9.650 & 1414 & 46.6 & 70.1 & 1.50 & 0.3210   & 44.237  & 7.33     \\
035423.25-062819.2 & 58.597  & -6.472 & 2292 & 57.0 & 84.7 & 1.49 & 0.4225   & 44.579  & 7.92     \\
084936.25+244905.4 & 132.401 & 24.818 & 2353 & 46.2 & 68.0 & 1.47 & 0.3527   & 44.601  & 7.95     \\
100541.86+433240.4 & 151.424 & 43.545 & 2320 & 32.2 & 53.2 & 1.65 & 0.1785   & 44.594  & 7.94     \\
103040.81+191406.6 & 157.670 & 19.235 & 3168 & 80.0 & 107.7 & 1.35 & 0.4686   & 44.836  & 8.33     \\
105053.78+344338.6 & 162.724 & 34.727 & 2671 & 87.4 & 137.8 & 1.56 & 0.1471   & 44.074  & 7.80     \\
105234.19+233902.5 & 163.142 & 23.651 & 3501 & 37.7 & 84.8 & 2.25 & 0.4964   & 44.949  & 8.47     \\
110036.57+064121.3 & 165.152 & 6.689  & 1200 & 15.0 & 38.9 & 2.58 & 0.2993   & 44.179  & 7.16     \\
111123.80+505131.0 & 167.849 & 50.859 & 3628 & 36.4 & 76.4 & 2.10 & 0.5595   & 45.034  & 8.55     \\
120700.30-021927.1 & 181.751 & -2.324 & 1359 & 44.0 & 60.1 & 1.37  & 0.3085   & 44.359  & 7.36     \\
120950.85+554113.1 & 182.462 & 55.687 & 5299 & 44.1 & 103.5 & 2.35 & 0.3390   & 44.497   & 8.61     \\
121549.43+544223.9 & 183.956 & 54.707 & 1359 & 26.3 & 46.2 & 1.75 & 0.1500   & 44.302   & 7.33     \\
125100.44+660326.8 & 192.752 & 66.057 & 1478 & 49.8 & 67.0 & 1.34  & 0.2820   & 44.910   & 7.71     \\
125343.71+122721.5 & 193.432 & 12.456 & 8516 & 74.7 & 191.1 & 2.56 & 0.2071   & 44.284   & 8.91     \\
130112.91+590206.6 & 195.304 & 59.035 & 3919 & 51.0 & 95.5 & 1.87 & 0.4764   & 45.761   & 8.98     \\
131411.15+083759.8 & 198.546 & 8.633  & 4499 & 45.2 & 96.0 & 2.12 & 0.3589   & 44.483   & 8.46     \\
133005.71+254243.7 & 202.524 & 25.712 & 7479 & 148.7 & 286.7 & 1.93 & 0.5975   & 44.897   & 9.11     \\
134704.91+144137.6 & 206.770 & 14.694 & 1845 & 19.8 & 38.8 & 1.96 & 0.1346   & 44.349   & 7.62     \\
141700.82+445606.3 & 214.253 & 44.935 & 2918 & 72.1 & 96.5 & 1.34 & 0.1135   & 44.173   & 7.93     \\
141956.71+373912.8 & 214.986 & 37.654 & 6143 & 48.3 & 145.2 & 3.01 & 0.4768   & 44.744   & 8.86     \\
144645.93+403505.7 & 221.691 & 40.585 & 2844 & 76.7 & 109.6 & 1.43 & 0.2673   & 45.152   & 8.39     \\
150245.36+405437.2 & 225.689 & 40.910 & 1202 & 20.2 & 42.8 & 2.12 & 0.2329   & 44.101   & 7.12     \\
151525.53+344440.8 & 228.856 & 34.745 & 3060 & 61.5 & 84.6 & 1.38 & 0.3784   & 44.170   & 7.97     \\
152939.29+203906.8 & 232.414 & 20.652 & 5758 & 47.8 & 84.8 & 1.77 & 0.1506   & 44.038   & 8.45     \\
165252.67+265001.9 & 253.219 & 26.834 & 3345 & 87.6 & 224.9 & 2.57 & 0.3501   & 44.937   & 8.43     \\
213026.74-070320.6 & 322.611 & -7.056 & 4173 & 40.0 & 60.2 & 1.51 & 0.1841   & 44.115   & 8.21     \\
\hline
\end{tabular} 
\end{table*} 

We model the continuum emission of the selected AGN taking into account the contribution from the accretion disk as well as from the 
 host galaxy starlight. The spectroscopic fibers of the SDSS have an aperture of 3'', so for the considered redshift range ($0.11\leq z \leq 0.60$) the 
light is collected from the region of diameter from 7 to 35 kpc. Thus we model in detail the spectral region of the H$\beta$ emission line 
with starlight included. This is a complex multi-parameter task so we perform this fitting in a few separate steps.

\subsection{Preliminary continuum fitting}
\label{sect:preliminary}

We first use our own Fortran code to fit the continuum without the starlight. The emission lines are masked during the fitting, 
with 27 regions excluded (for the list of the masked regions, see Table~\ref{tab:maska}). 
We model the disk emission assuming a power law shape, with a fixed power-law spectral index
(i.e., the flux is modeled as $F_{\lambda} \propto \lambda^{-7/3}$) corresponding to the simplest model of an accretion disk 
(Shakura \& Sunyaev 1973). We also test the possible effect of the curvature which is expected from the disk emission due to the presence of the 
maximum disk temperature, $T_{max}$, and in this case we model the disk spectrum as
\begin{equation}
F_{\lambda} \propto \lambda^{-7/3} exp(-h\nu/kT_{max}),
\end{equation}
where $h$ is the Planck constant and $k$ is the Boltzmann constant. 

The Fe II pseudo-continuum is modeled using the theoretical templates of Bruhweiler \& Verner (2008). The templates are named as in Bruhweiler \& Verner (2008), so the name specifies the local density, the ionization parameter, and the turbulent velocity. The advantage of this approach over the observational templates is that the Fe II shape is well modeled also under the intense emission lines. On the other hand, the observational templates may also include other pseudo-continuum contributions like Fe III, while the theoretical templates are limited to Fe II itself.
Fe III is known to contribute to UV parts of AGN spectra (e.g., Vestergaard \& Wilkes 2001), 
and some of the Fe III lines can also be prominent in the optical band, most likely 4658 \AA, 5270 \AA, 5011 \AA, and 4702 \AA~ 
(see the full list of lines in Keenan et al. 1992). However, strong Fe III in the optical spectra of AGN are usually not reported. We thus generally use one of the theoretical templates (d11-m20-20; this is the template for the particle density $10^{11}$ cm$^{-3}$, ionization parameter $10^{20}$ cm$^{-2}$ s$^{-1}$ and the turbulent velocity of 20 km s$^{-1}$) but we also test other templates from the same set as well as an observational template by Boroson et al. (2002) which was used by Shen et al. (2011). We also include the Balmer Continuum (BC), described as in Dietrich et al. (2002) and the references therein. Blue-ward of the Balmer edge ($\lambda \approx 3675$\,\AA), we use the Planck function $B_{\nu}(T_{e})$ (Grandi 1982) with a constant electron
temperature of 15\,000 K. The change of the optical depth with wavelength is computed using a simple formula:
\begin{equation}
\tau_{\nu} = \tau_{BE} \left(\frac{\nu}{\nu_{BE}}\right)^{-3},
\end{equation}
where $\tau_{BE}$ is the optical depth at the Balmer edge radiation frequency ($\nu_{BE}$).
Red-ward of the Balmer edge, blended hydrogen emission lines are generated using atomic data of Storey \& Hummer
(1995) with the recombination line intensities for case B (opaque nebula), $T_{e}$ = 15\,000 K, $n_{e} = 10^{8}-10^{10}$\, 
${\rm cm}^{-3}$.
The accounted Balmer emission lines covered excitation top level for the transitions in the range $10 \leq n \leq 50$.
  This method is basically similar to the method of Kovacevic et al. (2014) but they use even higher-order transitions 
  (up to 400) arguing that higher-order lines are necessary to effectively represent the flux at 3646 \AA ; this is 
  only important for modelling of very high quality of the data. 
  They also fit the Balmer component together with H$\beta$ which is not convenient in our case since we model the wide range continuum separately, 
  and later we model H$\beta$ only in the narrow wavelength range where the Balmer component is not present.

Fitting the continuum in this way we obtain the normalizations of the Fe II and BC pseudo-continuum for each object.

\subsection{Starlight fitting}
\label{sect:starlight}

We subtract the Fe II and BC pseudo-continua on the basis of the preliminary fits described above, and then we model the starlight contribution to the spectrum. We use the code STARLIGHT\footnote{http://astro.ufsc.br/starlight/node/1}, described by Cid Fernandes et al. (2005,2009), in the version based on 45 stellar templates of Bruzual \& Charlot (2003) corresponding to different stellar ages (15 options) and metallicities (3 options). STARLIGHT
also allows for the presence of a power law, and we fix again its slope at 7/3, as before; the relative normalization is arbitrary. We use the same mask as in Sect.~\ref{sect:preliminary} in order to avoid the bands strongly contaminated by AGN emission lines. The code allows for stellar extinction
. The code allows us to obtain the relative contribution of  45 stellar components and of a power law through a Monte Carlo Markov Chain approach. The method minimizes the number of requested components. The code also returns the stellar velocity dispersion. 

\begin{figure}
 \centering
  \includegraphics[width=0.95\hsize]{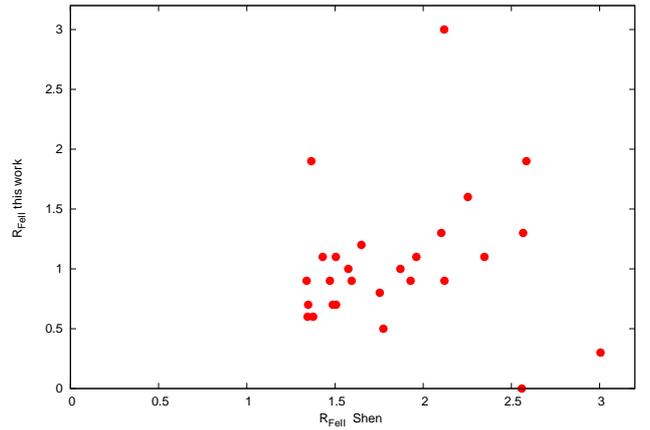}
 \caption{The comparison of the $R_{Fe}$ values in the Shen et al. catalog and from the new fits presented in this paper.}
 \label{fig:my_Shen}
\end{figure}

\begin{figure}
 \centering
 \includegraphics[width=0.95\hsize]{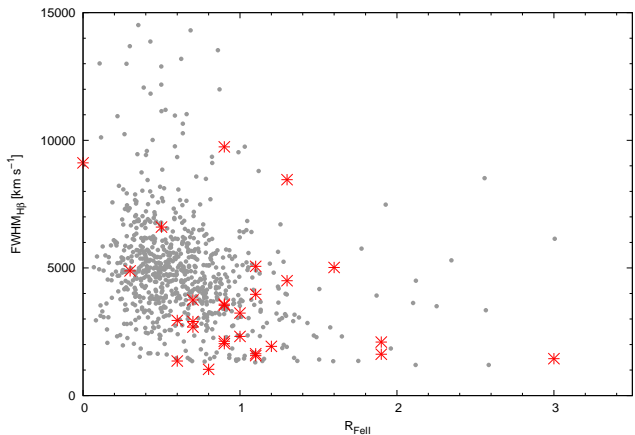}
 \caption{Location of the sources with measurement errors below 20\% in the Shen et al. (2011) catalog on the diagram showing the quasar main sequence, 
 with the $H _{\beta}$ line width plotted against the $R_{Fe}$ , measuring Fe II 
 strength (gray points). Red stars show the new location for the 27 objects fitted in this paper.}
 \label{fig:new_diagram}
\end{figure}

\subsection{Emission-line fitting and determination of a new value of $R_{Fe}$}
\label{sect:line_fitting}

Next we concentrate on the wavelength region 4400 - 5100 \AA~ where optical Fe II and H$\beta$ are calculated. Since this range
 is relatively narrow we fix the  shape of the starlight component as derived in Sect.~\ref{sect:starlight} as well as the starlight reddening. 
 However, we allow for changes in the normalization of the starlight, the power law contribution, and  Fe II.
 We now remove the use of the emission line mask and we model the line emission. The possibility of the extinction affecting the power 
 law component is also included. We use again the Cardelli extinction curve for that purpose as well, for simplicity.
 This prescription might not be good for intrinsic extinction in AGN but the differences between various extinction curves only become strongly apparent 
 in the UV part of the spectrum. We discuss this issue further below for one of the objects. 

H$\beta$ is modeled taking into account the decomposition into the broad and narrow component. Here we follow the approach 
of Shen et al. and tie the width of the narrow component of H$\beta$ to the [OIII] line, and both are represented by Gaussian lines.
The broad component of H$\beta$ is modeled either as a Gaussian or as a Lorentzian component. We also include broad H$\gamma$ 
in our fits since the tail of this line affects the spectrum close to 4400 \AA.  We neglected the presence of the He II 4686 since 
the visual inspection of the spectra did not indicate strong contribution from this line. Mean spectra usually do not show 
a clear presence of this line while in rms spectra the line is clearly seen (e.g., Peterson et al. 2014, Fausnaugh et al. 2017). 
With a single spectrum we cannot obtain a unique decomposition with this line included.
The [OIII] doublet is modeled assuming the theoretical ratio (1:3) between the strengths of the two components (Dimitrijevic et al. 2007).

The normalization of the Fe II component is again allowed to vary at this stage of the fitting, and we use the same templates as during the preliminary fits (see Sect.~\ref{sect:preliminary}).
The equivalent width of the Fe II is calculated in the 4434 - 4684 \AA~ range following the definition of Shen \& Ho (2014). 
We calculate the line EW with respect to the underlying power law, both for the Fe II and H$\beta$ broad components, that is, we do not include the starlight contribution. This results in a large modification in comparison with Shen et al. (2011) 
who do not differentiate between the accretion disk continuum and the starlight.

\begin{figure*}
 \centering
 \includegraphics[width=0.95\hsize]{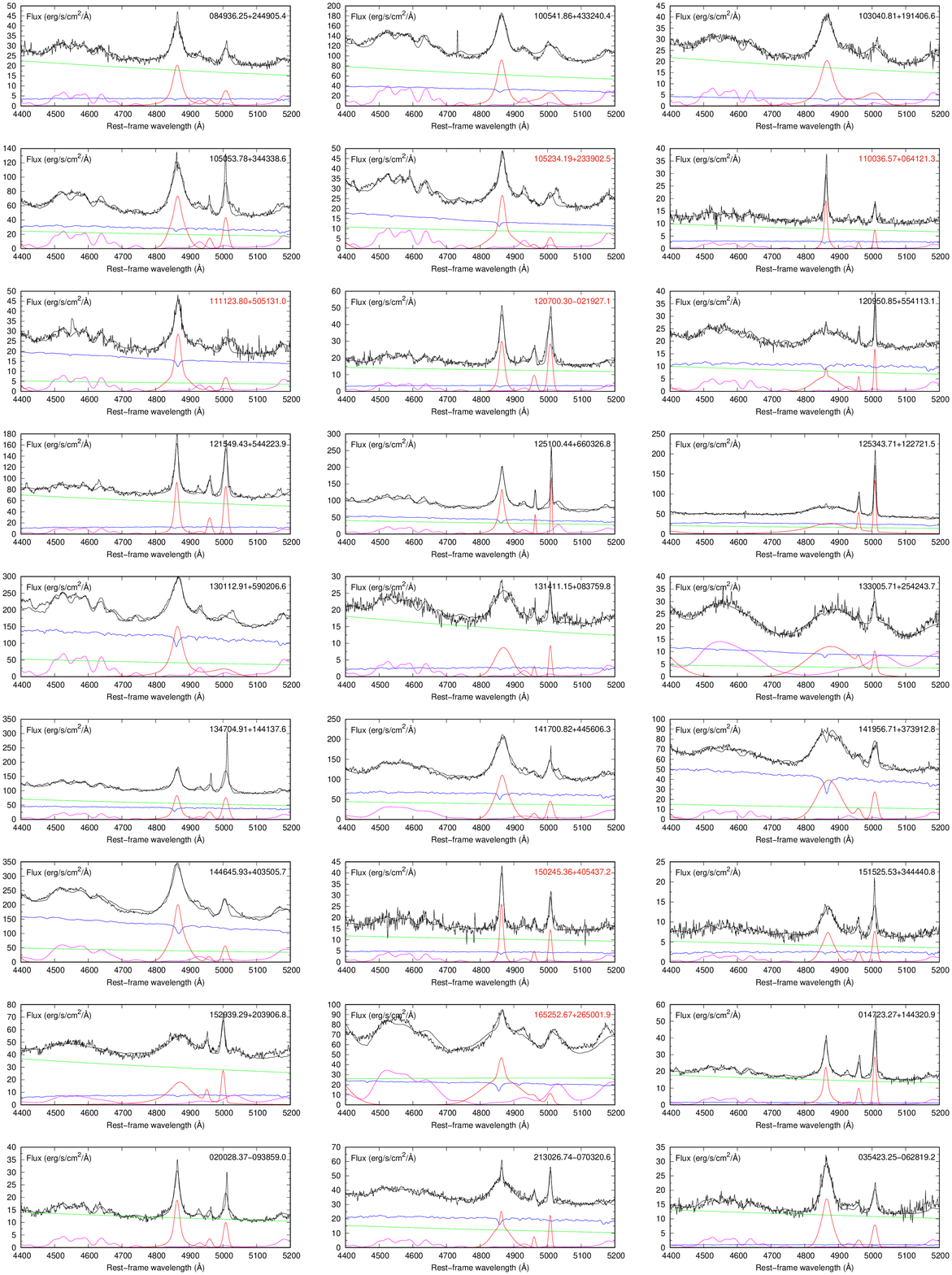}
 \caption{The region of the H$\beta$ line for all sources, with a black line showing the data and the best fit. The fit components are given in the following colors: emission lines (red), Fe II (magenta), disk power law including extinction (green), and starlight (blue). Objects with extreme values of $R_{Fe}$ after refitting have red labels.
}
    \label{fig:wszystkie}
\end{figure*}

\begin{table}
\caption{Masked regions during the preliminary continuum fitting (see Sect.~\ref{sect:preliminary}) and with STARLIGHT fitting.}   
\label{tab:maska}      
\centering                          
\begin{tabular}{l r r }        
\hline\hline      
No.    & $\lambda_{min}$  &  $\lambda_{max}$    \\
         & [\AA]    &  [\AA]  \\
\hline
1 & 1000.0  & 1565.0 \\  
2 & 1620.0  & 1680.0 \\
3 & 1880.0  & 1940.0 \\
4 & 2290.0  & 2340.0 \\
5 & 2390.0  & 2430.0 \\
6 & 2640.0  & 2660.0 \\
7 & 2780.0  & 2820.0 \\
8 & 3120.0  & 3140.0 \\
9 & 3160.0  & 3660.0 \\
10 & 3720.0 & 3740.0 \\
11 & 3850.0 & 3890.0 \\
12 & 3960.0 & 3980.0 \\
13 & 4080.0 & 4110.0 \\
14 & 4320.0 & 4360.0 \\
15 & 4670.0 & 4700.0 \\
16 & 4830.0 & 4900.0 \\
17 & 4940.0 & 4970.0 \\
18 & 4990.0 & 5030.0 \\
19 & 5400.0 & 5440.0 \\
20 & 5840.0 & 5940.0 \\
21 & 6280.0 & 6400.0 \\
22 & 6500.0 & 6620.0 \\
23 & 6700.0 & 6740.0 \\
24 & 7040.0 & 7080.0 \\
25 & 7840.0 & 7900.0 \\
26 & 8440.0 & 8470.0 \\
27 & 8830.0 & 9000.0  \\
\hline 
\end{tabular}
\end{table}

\section{Results}

Our analysis of the extreme EV1 sources shows that the careful individual approach to fitting their spectra is indeed important and modifies
the source parameters considerably.  For all sources initially selected from the Shen et al. (2011) catalog with the criterion $R_{Fe} > 1.3 $ 
and measurement errors smaller than 20\%, we derived new values of the line and continuum properties. In general, new values of $R_{Fe}$ are significantly 
smaller than the values given in the Shen et al. catalog. The results are given in  Table~\ref{tab:nasze} 
for both the Gaussian and Lorentzian shape of the H$\beta$ line. In 10
out of 27 cases, the Lorentzian fit was favored over Gaussian, 
so there was no clear pattern in the selected sources to favor Lorentzian shape, as is frequently the case for Narrow Line Seyfer 1 (NLS1) galaxies (Laor et al. 1999). 
None of the 27  sources have now  $R_{Fe} > 1.3$ for Lorentzian fits, and 9 sources have  $R_{Fe} > 1.3$ for Gaussian fits.
Considering only the better fits (Gaussian or Lorentzian) we have only 6 such sources. We plotted all new values against the old values 
in Fig.~\ref{fig:my_Shen}. The trend seems to some extent systematic, and we discuss the possible reasons below.
We show the new location of the selected extreme EV1 AGN in the $R_{Fe}$ versus FWHM diagram (see Fig.~\ref{fig:new_diagram}). 
Most objects simply moved towards the main cluster of points. Thus, in our analysis the objects with truly extreme EV1 
properties seem to be relatively rare. 

 Since our fitting procedures include starlight contamination and power-law extinction, we first summarize our findings below, 
 and we address the importance of these two aspects in data fitting.

\subsection{Starlight contamination}

Sources selected at the extreme tail of the EV1 are thought to have high Eddington ratios but at the same time they also have low values 
of the black hole masses (Shen \& Ho 2014), therefore their nuclear luminosity does not always outshine the rest of the galaxy.
The starlight contamination as determined from our data fitting ranges from 1.4 \% up to 77.8 \%  (see Table~\ref{tab:nasze}). 
The highest starlight contamination is in the source SDSS133005.71+590206.6.
which is notified for starburst activity in SDSS standard output\footnote{http://skyserver.sdss.org}. 

It is interesting to note that in general the stellar populations contaminating our sources consist of different stellar populations. 
 In Table ~\ref{tab:nasze} we give the age of the dominant population, and the most frequently are very young 1 Myr stars. 
In a few sources, however, the oldest population dominates.   In some objects an intermediate-age stellar population dominates, 
with ages between 10 and 100 Myr. These latter sources have relatively high contamination from starlight but there is no unique relation between the 
contamination level and the stellar population age. This is most likely related to the episodic nature of AGN activity (e.g., Czerny et al. 2009). 
The velocity dispersion coming out from the starlight fits is typically around 165 km s$^{-1}$. Taking into account the spectral resolution of 
SDSS data and that of the STARLIGHT models gives an average value of 175 km s$^{-1}$ (see e.g., Bian \& Huang 2010). 
The extinction in the starlight is not determined reliably since it is strongly coupled to the possible extinction towards 
the accretion disk which is neglected in the STARLIGHT code. However, this is not essential in the analysis of the H$\beta$ line region.

\subsection{Power-law extinction}

 Optical data is usually fitted with an accretion disk component approximated as a power law with an index left as a free parameter. In our fitting we fixed this index according to prediction of the standard Shakura-Sunyaev disk since the standard disk model was shown to be a good description of the disk continuum (e.g., Czerny et al. 2011, Capellupo et al. 2015) but we allowed for an extinction which affects the slope. We used the same extinction curve (Cardelli law) for the power law component and for starlight, for simplicity. The extinction fitted ranges from 0 to 1.73, but with mean and median values of 0.30 and 0.13, respectively. So for most sources, apart from object SDSS165252.67+2650001.9, the extinction is relatively moderate. For this latter, extreme object, we tried another extinction law, more appropriate to quasars (Czerny et al. 2004). This extinction curve has no 2175 \AA~ bump, and is generally flatter than the Cardelli law. However, the shape of the extinction curve is not essential when extinction is moderate, and only the spectral region 4400 - 5100 \AA~ is considered, as in our approach at the stage of emission line fitting. It could have significantly more importance, particularly for the starlight composition, during the first-stage, wide-range fitting, but for now the power law extinction this is not taken into consideration. We checked the effect of the extinction law change in the case of SDSS165252.67+2650001.9. With the new extinction law we obtained a statistically much better fit ($\chi^2/dof$ dropped from 5.46 down to 3.33 but the values of the key line parameters did not change, and the new value of $R_{Fe}$ was 2.05).

\subsection{H$\beta$ properties}

A Gaussian shape is most frequently used for modeling Seyfert 1 galaxies, and the Lorentzian shape is considered to be more suitable for 
the NLS1 (Laor et al. 1997). 
In our sample we have eight sources with a line width in the Shen et al. catalog below 2000 km s$^{-1}$, while the rest of objects have broader lines.
We thus used both Gaussian and Lorentzian shapes for the broad H$\beta$ component. A Lorentzian shape was better in roughly half of the sample,
including some objects with broad emission lines. In the case of quasars, the usual limit between type A and type B sources (see Sulentic et al. 2000) is at 4000 km s$^{-1}$ 
instead of 2000 km s$^{-1}$ (with such a criterion, we would have only 5 sources with broad lines) 
but the selected sources are not of the class of distant bright quasars, with the 5100 \AA~monochromatic flux in the Seyfert galaxy class, 
below $10^{45}$ erg s$^{-1}$. Formally our sources are located in A3 and A4 parts of the quasar classification diagram of Suletic et al. (2002) 
but they are fainter than the sources considered in this paper (see e.g., Zamfir et al. 2010 for the mean values in their subsamples). Therefore, for our sources the standard classification based on the limit of 2000 km s$^{-1}$ should apply. Nevertheless, for all six extreme sources Gaussian fits are actually better, independently from the differences in the line width, although in the three sources with narrower lines the difference between the Gaussian and Lorentzian fits are quite small.

 Lines are symmetric; we do not see any significant asymmetry which is sometimes present in very bright quasars with broad emission lines 
 (e.g., Zamfir et al. 2010; Rakic et al. 2017). Most of them are best fitted when the narrow component (fitted as a Gaussian) 
 is included; although in most objects this component is not well separated visually. 
  Thus, decomposition of H$\beta$ may be a considerable source of errors.

 In our fits we assume that the width of the narrow H$\beta$ component is equal to the width 
 of [OIII]5007\AA~which allowed us to obtain a relatively unique solution for the line decomposition. 
 The contribution of the narrow component to the total line EW is not very high, with a mean value of the ratio of the narrow line component 
 to the broad line component of 0.25, and a median of only 0.12. This is generally consistent with a weak narrow line contribution to
 H$\beta$ even in the large sample of NLS1 (Zhou et al. 2006). 
 Thus, if $R_{Fe}$ is measured with respect to the total H$\beta$ flux, the mean value in our sample drops from 1.10 to 0.84. 
 Thus, for most sources the determination of the narrow component contribution is not a key issue but for a few objects it is indeed important.
 The highest narrow component contribution (narrow to broad component ratio ratio of 1.24) is seen in SDSS150245.36+405437.2 which has the highest
 $R_{Fe}$ value in our sample. If $R_{Fe}$ is measured with respect to the total H$\beta$ flux, the new value is 1.4. This object  would still, therefore, have 
 been selected as an extreme EV1 case.

\begin{figure}
 \centering
 \includegraphics[width=0.95\hsize]{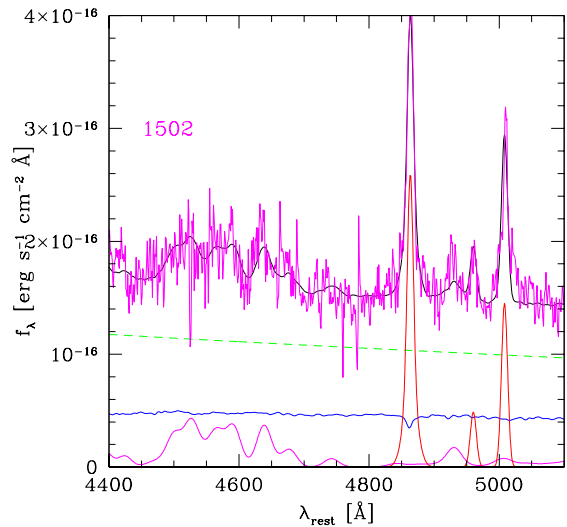}
 \caption{The expanded region of the H$\beta$ line of the source SDSS150245.36+405437.2 (magenta line) which shows an extremely narrow line and the highest value of $R_{Fe}$. Fit components: total flux (black line), starlight (blue line), power law (green dashed
 line), and total line contribution (red line). The lower magenta line shows the shape of the Fe II template d11-5-m20-20-5-mod broadened with a Gaussian of 700 km s$^{-1}$.}
 \label{fig:narrow_line}
\end{figure}

\begin{figure}
 \centering
 \includegraphics[width=0.95\hsize]{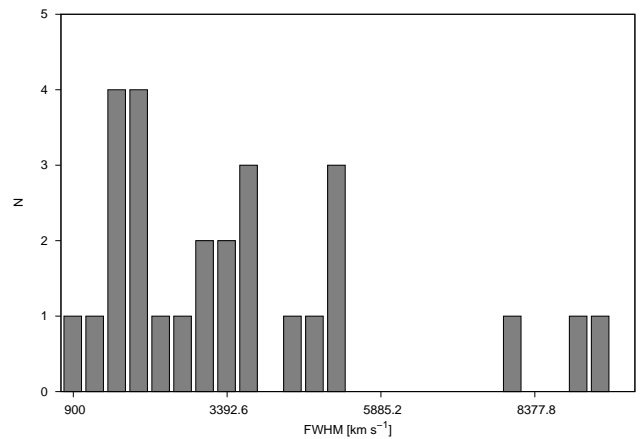}
 \caption{The distribution of the line widths in our sample of 27 objects.}
\label{fig:FWHM}
\end{figure}

  Some of the objects, such as SDSS121549.43+544223.9, or SDSS150245.36+405437.2 as mentioned before,
 have a very narrow H$\beta$ line, but their kinematic width is still somewhat broader than the width of  [OIII]
  (870 km s$^{-1}$ and 665  km s$^{-1}$, respectively). Intensities of the [OIII] lines are 
 low, so the objects meet the criterion [OIII]/H$\beta <$ 3 (this ratio for the broad H$\beta$ component only is 1.85 and 1.29, respectively, for these two objects
 in our fits). The contribution of the disk emission is important, so the source cannot be considered as a Seyfert 2 galaxy. 
 The relative shift between [OIII] and H$\beta$ is less than 45  km s$^{-1}$, for both sources. 
 The objects are well fitted (see Fig.~\ref{fig:wszystkie} for all fitted spectra and Fig.~\ref{fig:narrow_line} for an expanded version 
 for SDSS150245.36+405437.2), with noticeable Fe II. The distribution in the broad line H$\beta$ FWHM is shown in Fig.~\ref{fig:FWHM}.

\subsection{Extreme $R_{Fe}$ objects}

\begin{figure}
 \centering
  \includegraphics[width=0.95\hsize]{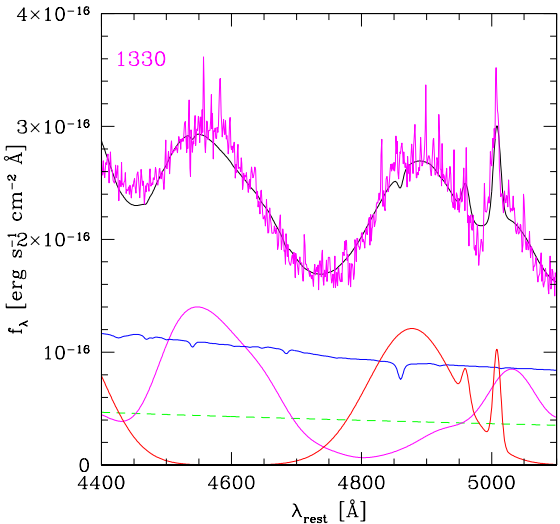}
 \caption{The expanded region of the H$\beta$ line of the source SDSS133005.71+254243.7  
   (magenta line) which shows an extremely broad H$\beta$ and at the same time very strong and broadened Fe II emission.
   Fit components: total flux (black line), starlight (blue line), power law (green dashed line), total line contribution (red line), 
 and lower magenta shows the shape of the Fe II template d11-m10-20-5 broadened with a Gaussian of 2300 km s$^{-1}$.}
    \label{fig:broad_line}
\end{figure}

In most cases the value of the $R_{Fe}$ derived from our analysis is smaller than originally derived by Shen \& Ho (2014). 
However, six objects still satisfy our initial criterion of $R_{Fe} > 1.3$, and they now form our class of extreme EV1 objects
(see Table~\ref{tab:nasze}). 
 With only six sources left representative of extreme EV1,  statistics are unclear, nevertheless they do not seem
to form a homogeneous class. Three of them are NLS1 objects, with FWHM narrower than 2100 km s$^{-1}$, and two of those have extremely narrow lines.
However, the remaining three objects have very broad lines, above 5400 km s$^{-1}$, so even in the quasar classification they belong to class B, 
and not class A, since the limit for massive quasars separating low and high Eddington ratio is at 4000 km $^{-1}$ (Sulentic et al. 2000). 
They populate the region which is not considered by Zamfir et al. (2010) to contain  extreme objects, and were intentionally left out of their SDSS sample.

Equivalent width  (EW) of H$\beta$ in our whole sample, if measured with respect to the absorbed power law as performed in our approach, is quite typical 
(see Zhou et al. 2006 for NLS1 and Forster et al. 2001 for bright quasars). In the whole Shen et al. catalog the mean EW(H$\beta$) = 82.9 \AA. 
The three extreme broad line sources have a line strength of that order, while the three extreme narrow-line sources have EW(H$\beta$) 
from 17.2 to 26.8 \AA, and they are in the tail of the distribution (see Zhou et al. 2006). In two of those objects
reddening of the power law component is relatively large ($A_V = 1.31$), so if we measure H$\beta$ with respect to uncorrected power law the
broad line is very faint (EW(H$\beta$) = 11.1 \AA \ and 8.1 \AA, instead of 26.8 \AA \ and 17.2 \AA). In such a case   these two  sources
would have been classified as a Weak Line Quasar (Wu et al. 2012). However, if the narrow component is included, line width would go up to 22.4 and 18.1 \AA,
correspondingly. Therefore, independent of the details, line strength in these three objects is low.
The reddening of course does not affect the measurement of $R_{Fe}$.

\begin{figure*}
 \centering
 \includegraphics[width=0.95\hsize]{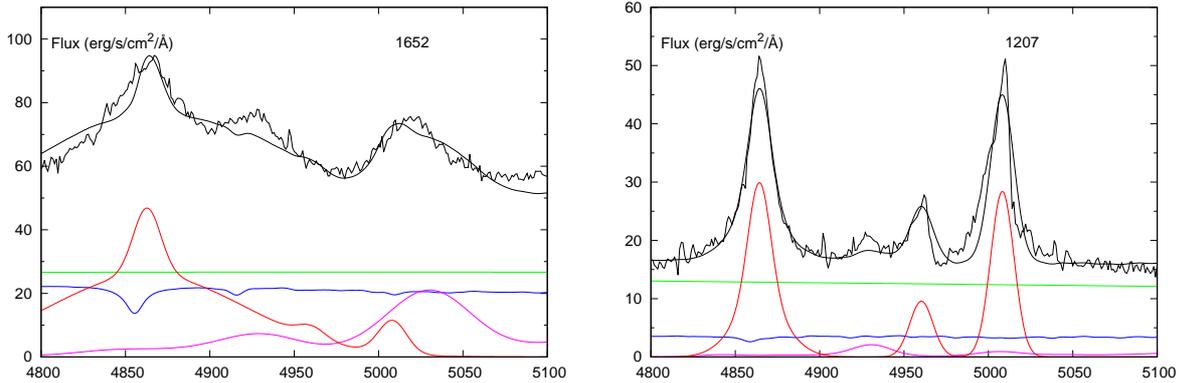}
 \caption{The expanded region of the OIII line of the sources SDSS165252.67+265001.9 (left panel) and SDSS120700.30-021927.1 (right panel). Line colors as in Fig.~\ref{fig:wszystkie}.}
    \label{fig:OIII_region}
\end{figure*}

In their studies, Shen \& Ho (2014) nicely illustrated the systematic trend in the EW([OIII]) 
line with $R_{Fe}$ ;  we therefore also checked whether our extreme objects follow this pattern. However, the [OIII] emission seems to be strong in all six objects, 
with EW([OIII]) from  11 to 43 \AA, much higher than expected from their diagram. On the other hand, there is indeed no systematic difference 
between our three narrow- and three broad-line objects, as stressed by Shen \& Ho (2014).
We give the old and new values of  EW([OIII]) for the six most extreme sources in Table~\ref{tab:OIII},
and we show the expanded version of the fitted spectra for two objects in Fig.~\ref{fig:OIII_region}. 
The decomposition in the [OIII] region is difficult since there are two strong Fe II transitions 
(at 4930.70 \AA~ and 5031.87 \AA). In our fits the normalizations of these lines come from the template fitting 
(a single normalization of the entire Fe II pseudo-continuum), which might not be correct (see Kovacevic et al. 2010). 
The issue of the spectral fitting in an automatic mode was already raised in Sulentic \& Marziani (2015), and the current paper shows it clearly.

\begin{figure*}
 \centering
 \includegraphics[width=0.95\hsize]{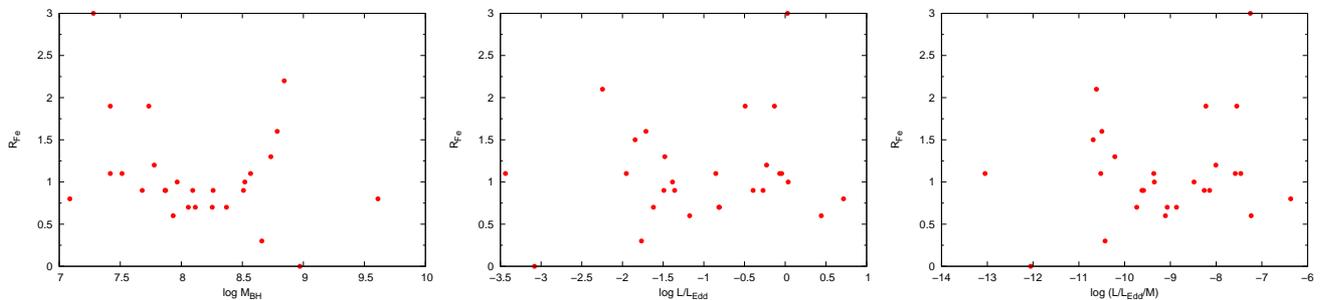}
 \caption{The relation between $R_{Fe}$ and the black hole mass (left), with the Eddington ratio (middle) and the hypothetical driver of EV1 representing the maximum of the accretion disk temperature (right panel). Green points illustrate extreme objects.}
    \label{fig:trzy}
\end{figure*}

\subsection{Black hole mass and Eddington ratios in our sample}
\label{sect:mass}

Since now our fits cover, in practice, the whole parameter range, and not just the extreme objects, we checked whether or not in our sub-sample we could see a direct relation between the  $R_{Fe}$ and the black hole mass or Eddington ratio. We thus obtained the values of the black hole masses based on our new determination of the line width, using the standard single spectrum method. We used the exact coefficients from Bentz et al. (2013), from their fit {\it Clean}. We did not introduce a correction to the monochromatic luminosity for starlight content since the starlight contamination was not very high in most of our objects. The bolometric luminosity was derived from the Shakura-Sunyaev disk model using the derived black hole mass,  5100 \AA \ luminosity, and adopting 10\% efficiency (see Tripp et al. 1994).  The results are given in Table~\ref{tab:summary}, and plotted in Fig.~\ref{fig:trzy}, left and middle panel. The correlations are not significant, the corresponding values of the correlation parameters are -0.29 and +0.30.
Our extreme EV1 objects do not occupy a special region of that diagram. Three narrow line objects have small masses, just above $10^7 M_{\odot}$, and high Eddington ratios close to 1, while the three broad line objects have masses approaching $10^9 M_{\odot}$, and Eddington ratios of a few per cent.

 The dispersion in the H$\beta$ line width in the Shen et al. (2011) catalog for objects with $R_{Fe} > 1.3$ is also large, 
 and our new fitting did not remove it, but simply reduced it slightly. Taking into account that the luminosity range in $\log L_{5100}$ 
 is quite narrow points out that the Eddington ratio cannot be the sole driver of the EV1.

 We  also considered the possibility  that the EV1 is driven not just by the Eddington ratio but by the position of the maximum of the wide-range quasar spectrum (Panda et al., 2017). In the case of the simplest accretion-disk model (Shakura \& Sunyaev 1973) this maximum is connected with the maximum of the accretion disk temperature, and it scales with the black hole mass and accretion rate as 
\begin{equation}
\nu_{max} \propto \bigg({\dot M  \over M_{BH}^2}\bigg)^{1/4}, 
\end{equation}
that is, $\nu_{max} \propto (\dot m /M_{BH})^{1/4}$ (Tripp et al. 1994). 
 We thus additionally plotted the ratio of the Eddington ratio to the black hole mass in order to see whether or not our extreme sources cover 
a similar range in this parameter (see Fig.~\ref{fig:trzy}, right panel). However, this is not the case.

 We interpret this in the following way. The accretion rate, $\dot M$, is related to the bolometric luminosity, $L_{bol}$ ( $L_{bol} = \eta \dot M c^2$, where $\eta$ is the accretion efficiency), and the bolometric luminosity is usually estimated from the $L_{5100}$ luminosity ($L_{bol} = 9 L_{5100}$). The black hole mass in turn is determined again from the $L_{5100}$ luminosity, and from the FWHM of the line as in Bentz et al. (2013). When the two are combined, the dependence on the luminosity vanishes, and we obtain a simple relation:
\begin{equation}
\nu_{max} \propto FWHM^{-1}.
\end{equation}
 Therefore, broad-line and narrow-line objects do not cover the same region, and the hypothesis of the disk maximum temperature 
 as the driver is also not promising.

\begin{table*}[h]

\caption{Fits with the assumption of the Lorentzian and Gaussian shape for H$\beta$ broad component.}

\label{tab:nasze}

\begin{tabular}{@{}llllllll|llllll@{}}
\toprule
Name & Stellar  & Starlight  & EW(FE)  & EW(H${\beta}$)  & R$_{Fe}$   & FWHM &  $\chi^2$  & Starlight   & EW(FE)  & EW(H${\beta}$)  & R$_{Fe}$  & FWHM  &  $\chi^2$  \\
      & age {[}Gyr{]}       & {[}\%{]}      &      &           &      &  km s$^{-1}$    &  d.o.f.       & {[}\%{]}       &       &        &      &  km s$^{-1}$ & d.o.f.   \\ \midrule

0147 & 0.005                     & 7.5                & 34.8  & 37.5  & 0.9 & 2034   & 4.99  & 7.7          & 33.1       & 27.1       & 1.2     & 2664            & 5.12       \\
0200 & 13.000                    & 1.2                & 44.4  & 40.9  & 1.1 & 1566   & 3.56  & 1.3          & 41.8       & 30.3       & 1.4     & 2345            & 3.75       \\
0354 & 0.001                     & 5.8                & 46.3  & 76.4  & 0.6 & 2266   & 1.98  & 5.5          & 38.1       & 54.0       & 0.7     & 2684            & 1.94       \\
0849 & 0.025                     & 11.4               & 43.9  & 50.6  & 0.9 & 2128   & 1.53  & 1.3         & 41.5       & 37.3       & 1.1     & 2743            & 1.61       \\
1005 & 0.010                     & 30.0               & 73.6  & 63.2  & 1.2 & 1930   & 3.06  & 29.9         & 63.6       & 48.8       & 1.3     & 2732            & 3.79       \\
1030 & 0.003                     & 15.1                 & 53.2  & 79.6  & 0.7 & 2902   & 1.63  & 15.1         & 44.9       & 60.6       & 0.7     & 3690            & 1.83       \\
1050 & 0.001                     & 44.2               & 133.9 & 177.8 & 0.8 & 2708   & 4.77  & 44.8         & 142.6      & 147.0      & 1.0     & 3233            & 4.37       \\
1052 & 0.001                     & 50.3                 & 146.1 & 110.1 & 1.3 & 5392   & 5.11  & 51.1         & 136.0      & 87.2       & 1.6     & 5020            & 4.44       \\
1100 & 0.001                     & 21.6               & 42.9  & 28.4  & 1.5 & 1502   & 1.55  & 21.6         & 34.3       & 18.3       & 1.9     & 1620            & 1.48       \\
1111 & 0.001                     & 69.1               & 220.0 & 187.6 & 1.2 & 3828   & 3.43  & 64.7         & 220.2      & 171.5      & 1.3     & 4502            & 2.83       \\
1207 & 0.102                     & 10.9               & 43.1  & 33.9  & 1.3 & 1398   & 2.34  & 15.3         & 50.1       & 26.8       & 1.9     & 2099            & 2.30       \\
1209 & 0.001                     & 47.4               & 77.2  & 87.7  & 0.9 & 4212   & 2.36  & 43           & 72.0       & 68.2       & 1.1     & 5065            & 2.07       \\
1215 & 11.000                    & 11.7               & 24.2  & 28.7  & 0.8 & 1030   & 3.99  & 15           & 26.7       & 22.0       & 1.2     & 1555            & 4.39       \\
1251 & 0.010                     & 56.6               & 91.37  &  124.70& 0.7 & 1377   & 4.38  & 54.3       & 46.44      & 79.75      & 0.6     & 2025            & 8.71       \\
1253$^*$ & 0.001                 & 47.4               & 12.2  & 315.1 & 0.0 & 9118   & 3.66  & 50.7         & 10.5       & 210.8      & 0.0     & 8863            & 4.26       \\
1301 & 0.025                     & 65.6               & 190.0 & 196.2 & 1.0 & 2318   & 5.54  & 62.8         & 170.0      & 157.9      & 1.1     & 3517            & 6.45       \\
1314 & 2.500                     & 9.8                & 40.6  & 51.3  & 0.8 & 3124   & 2.09  & 9.8          & 36.5       & 39.2       & 0.9     & 3584            & 1.86       \\
1330 & 0.001                     & 92.4               & 975.2 & 1199.4& 0.8 & 9404   & 1.43  & 88.5         & 530.2      & 565.9      & 0.9     & 9742            & 1.24       \\
1347 & 0.102                     & 36.1               & 48.7  & 45.8  & 1.1 & 1640   & 8.32  & 36.3         & 43.7       & 34.8       & 1.3     & 2048            & 8.60       \\
1417 & 0.010                     & 35.9               & 83.5  & 132.2 & 0.6 & 3106   & 3.80  & 50.7         & 126.5      & 133.2      & 0.9     & 3537            & 3.41       \\
1419 & 0.286                     & 69.5               & 117.0 & 370.4 & 0.3 & 4258   & 2.60  & 67.8         & 91.2       & 280.9      & 0.3     & 4885            & 2.36       \\
1446 & 0.010                     & 78.2               & 307.9 & 374.5 & 0.8 & 2412   & 6.03  & 74.4         & 208.0      & 197.1      & 1.1     & 3968            & 5.89       \\
1502 & 0.010                     & 20.5               & 44.5  & 21.0  & 2.1 & 936    & 1.09  & 26.7         & 51.0       & 17.2       & 3.0     & 1445            & 1.02       \\
1515 & 0.001                     & 26.6               & 56.2  & 107.6 & 0.5 & 2816   & 1.38  & 26.6         & 49.9       & 78.3       & 0.6     & 2945            & 1.32       \\
1529 & 0.001                     & 20.8               & 34.5  & 72.3  & 0.5 & 5076   & 1.25  & 21.5         & 45.6       & 86.2       & 0.5     & 6603            & 1.16       \\
1652 & 0.025                     & 41.2               & 280.4 &  271.5& 1.0 & 9292   & 4.61  & 37.3         & 220.1      & 142.3      & 1.3     & 8457            & 4.36       \\
2130 & 0.010                     & 54.8                 & 66.6  & 107.9 & 0.6 & 3104   & 2.02  & 54.6         & 58.0       & 82.0       & 0.7     & 3746            & 1.44      
\end{tabular}

\vspace{1ex}

     \raggedright $^*$ The results for this object reported in the Shen et al. (2011) catalog were incorrect due to a software problem (Yue Shen, private communication).

  \label{tab:test}
\end{table*}

\begin{table}[]
\centering
\caption{Equivalent widths of [OIII] line for extreme objects from Shen et al. (2011) catalog and from our fitting.}
\label{tab:OIII}
\begin{tabular}{rrr}
\hline\hline
\ Name & EW(OIII)\_Shen & EW(OIII) this work \\
\  & \AA & \AA \\

\hline
1052   & 0              & 27.7  \\
1100   & 3.18           & 23.0  \\
1111   & 1.94           & 39.4  \\
1207   & 32.11          & 42.9  \\
1502   & 10.43          & 22.2  \\
1652   & 4.88           & 10.8 \\
\hline
\end{tabular}
\end{table}

\begin{table}[]
\centering
\caption{Summary of the results, selecting a Lorentzian or Gaussian fit for H$\beta$ as the basis of a better fit. }
\label{tab:summary}
\begin{tabular}{@{}lllll@{}}
\toprule
Name & logL$_{5100}$ & FWHM  & $R_{Fe}$ & log$M_{BH}$ \\ \midrule
0147 & 44.6896 & 2034 & 0.9 & 7.87 \\
0200 & 44.2369 & 1566 & 1.1 & 7.42 \\
0354 & 44.5791 & 2684 & 0.7 & 8.06 \\
0849 & 44.601 & 2128 & 0.9 & 7.87 \\
1005 & 44.5939 & 1930 & 1.2 & 7.78 \\
1030 & 44.8364 & 2902 & 0.7 & 8.25 \\
1050 & 44.0736 & 3233 & 1.0 & 7.97 \\
1052 & 44.9492 & 5020 & 1.6 & 8.79 \\
1100 & 44.1789 & 1620 & 1.9 & 7.42 \\
1111 & 45.0336 & 4502 & 1.3 & 8.73 \\
1207 & 44.3587 & 2099 & 1.9 & 7.73 \\
1209 & 44.4972 & 5065 & 1.1 & 8.57 \\
1215 & 44.3019 & 1030 & 0.8 & 7.09 \\
1251 & 44.9104 & 1356 & 0.6 & 7.42 \\
1253 & 44.284 & 9118 & 0.0 & 8.97 \\
1301 & 45.7612 & 2318 & 1.0 & 8.52 \\
1314 & 44.4832 & 3584 & 0.9 & 8.26 \\
1330 & 44.8972 & 9742 & 0.9 & 9.17 \\
1347 & 44.3492 & 1640 & 1.1 & 7.51 \\
1417 & 44.1726 & 3537 & 0.9 & 8.09 \\
1419 & 44.7441 & 4885 & 0.3 & 8.66 \\
1446 & 45.1518 & 3968 & 1.1 & 8.53 \\
1502 & 44.1005 & 1445 & 3.0 & 7.28 \\
1515 & 44.1699 & 2945 & 0.6 & 7.93 \\
1529 & 44.0382 & 6603 & 0.5 & 8.37 \\
1652 & 44.937 & 8457 & 1.3 & 9.07\\
2130 & 44.1152 & 3746 & 0.7 & 8.11 \\ \bottomrule
\end{tabular}
\end{table}

\section{Discussion}

Our analysis confirms the existence of the extreme EV1 quasars, although 
they seem rare when more accurate data fitting is adopted.  We find that the reliable measurement of the Fe II and the decomposition of the H$\beta$ 
line into broad and narrow components are difficult. This is seen already in the Shen et al. (2011) catalog. The mean and the median of the $R_{Fe}$ 
measuring the ratio of the Fe II to the broad H$\beta$ component are 0.97 and 0.70, respectively, in the whole sample, but when we limit the sample to measurements
with the errors below 20\%, those values drop to 0.64 and 0.38, respectively. The results from our fitting bring the $R_{Fe}$ values further down. 
Out of the 27 originally selected objects with $R_{Fe} > 1.3$, 6 objects remain in this class. However, our fitted sample has mean and median values $R_{Fe}$
of (excluding the object SDSS125343.71+127221.5) 1.02 and 0.90, respectively, meaning that they all statistically represent rather large EV1 values.

 The six most extreme Fe II emitters do not seem to form a homogeneous sample. Three of them have a narrow H$\beta$ line, out of them two can 
 be classified as NLS1s, and one narrowly fails to meet the formal criterion of 2000 km s$^{-1}$. The other three have very broad lines, above
 4500 km s$^{-1}$. These should belong to type B quasars (Sulentic et al. 2000) but they actually occupy the avoidance region as is defined 
 in Sulentic et al. (2002) and Zamfir et al. (2010). If the black hole mass is measured in a standard way, the first group have low masses and high 
 Eddington ratios (between 0.3 and 1) but the second group have high masses and low Eddington ratios (between 0.01 and 0.03). The Fe II broadening roughly 
 follows the H$\beta$ trend, that is, broad H$\beta$ sources require the Fe II broadening with a Gaussian of $\sim 2000$ km s$^{-1}$, while in the
 narrow H$\beta$ sources this broadening is of the order of $\sim 700$ km s$^{-1}$. So in our case the Fe II contribution is unlikely to come from the 
 surrounding starburst, as discussed by Lipari et al. (1993). The stellar content in the six objects is moderate, between 15 \% and 65 \%, so the starlight 
 contamination does not strongly affect the black hole mass determination; the effect is below 0.2 dex in the log scale.

 These results  may imply that the Eddington ratio itself is not the  sole driver of EV1. High Eddington-ratio sources monitored by
 Du et al. (2015, 2016) occupy the region between $R_{Fe} = 0.5$ and $R_{Fe} = 1.9$ (P. Du, private communication).
  In Panda et al. (2017) we considered the possibility that the maximum disk temperature is the most promising principal driver of the EV1 as it 
 determines the maximum of the SED, and the spectral shape of the incident  continuum is likely affecting the line ratios in BLR. 
 However, the two separate groups among our six sources also indirectly suggest a very different peak position and this explanation is also unlikely.

 The only property the two groups of sources have in common is the EW([OIII]5007), in the range of  22 to 45 \AA, apparently independent from the 
 H$\beta$ line width; this was implied by EV1 analysis by Boroson \& Green (1992) from the very beginning as the anticorrelation between
 $R_{Fe}$ and the peak of [OIII]5007 was the strongest effect. The nature of this anticorrelation thus still requires a more detailed physical explanation.

 Our results are complementary to the recent study of Cracco et al. (2016). They analyzed a large sample of NLS1 galaxies and confirmed the 
 generally Lorentzian shape of their Balmer lines but also considerable overlap in the properties with Seyfert 1 galaxies. 
 Some of their NLS1 objects did not have strong Fe II emission. We, in turn, analyzed some broad-line objects which do have strong Fe II emission. 
 Thus, strong Fe II emission and narrow lines are not as tightly related as previously thought.

 There is still a possibility that either our fits, or the method of black hole mass and Eddington ratio estimates, are not accurate enough. 
 We used the standard method for the black hole mass determination (see Sect.~\ref{sect:mass}),
 and the method outlined by Tripp et al. (1994) to get the bolometric luminosity. If the bolometric luminosity is calculated using the conversion 
 factor 9 between the bolometric luminosity and the 5100 \AA \ monochromatic luminosity, the gap between broad and narrow line objects is slightly 
 smaller but the systematic trend in the Eddington ratio still remains. Pennell et al. (2017) suggested determining the bolometric 
 luminosity using EW([OIII]) and $R_{Fe}$. This method gives slightly larger gaps in the Eddington ratio between the two groups of our extreme EV1 sample.

Our analysis takes into account the Fe II contamination much more carefully than the analysis done by 
Shen et al. (2011), and accounts for the starlight. However, we cannot yet achieve fully self-consistent fits. 
The STARLIGHT code does not allow the inclusion of power law reddening to the power law or the effects of the maximum of the disk temperature,
that is, the flattening of the power law at shorter wavelengths. 
There is still a possibility for further improvement in data fitting although it is not likely to change our calculations.

\subsection{Limitations of the current approach}

\begin{figure}
 \centering
 \includegraphics[width=0.95\hsize]{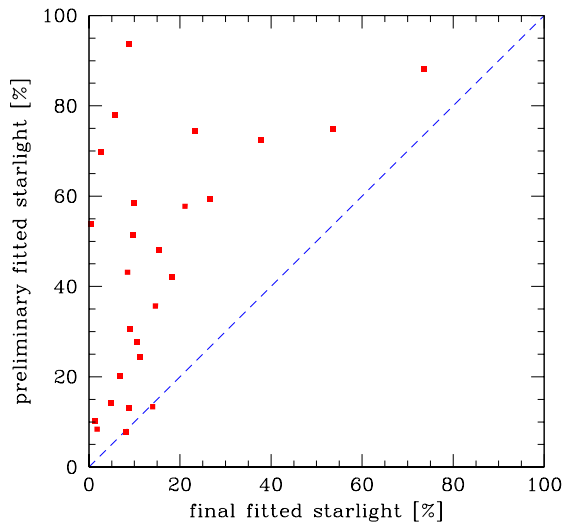}
 \caption{The relation between the amount of starlight determined in the preliminary fitting from STARLIGHT code and in the final fitting in narrower spectral band including emission lines and power law extinction.}
 \label{fig:starlight_change}
\end{figure}

\begin{figure}
 \centering
 \includegraphics[width=0.95\hsize]{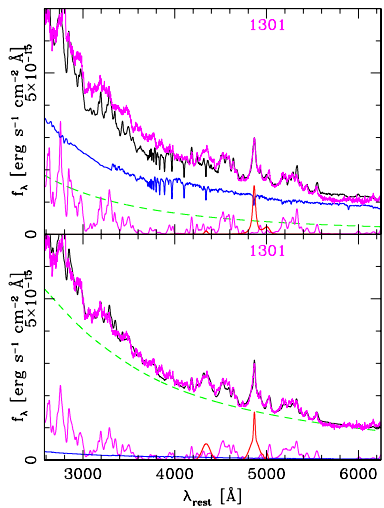}
 \caption{The wide-range fitting of the spectrum of SDSS130112.91+590206.6 without the automatic use of the STARLIGHT code (lower panel). The spectrum shows much less starlight contamination than the original results
based on automatic STARLIGHT run (upper panel) and further refinement in the 4400-5100 \AA~ spectral band.}
 \label{fig:broad_band}
\end{figure}

  Fitting the stellar content is important but high accuracy of the results is difficult to achieve.
  Statistical error might be small but the results strongly depend on the details of the adopted procedure.
  We fitted starlight in two stages, first for the wide-range data, using a power law with fixed slope and no intrinsic extinction (code STARLIGHT), 
  and later in the narrower 4400 - 5100 \AA~ band, with power-law extinction and all spectral features included (see Sect.~\ref{sect:line_fitting}). 
  We compared the values obtained at the two steps in Fig.~\ref{fig:starlight_change}. 
  The amount of starlight in general dropped in the last stage.
  This is a problem which cannot be solved easily. We cannot favor any of the obtained values.
  The first-stage fit is done using the wide-range data but without the extinction of the power law component since there is no such option 
  in the STARLIGHT code. The second-stage fitting takes this extinction into account but the fit is done in the narrower band, 
  without modification of the starlight content (relative proportions between the stellar populations).
  Our fitting code is not well adjusted to fit as many free parameters as STARLIGHT but nevertheless in one case a fitting was attempted. 
  We selected quasar SDSS130112.91+590206.6 for which the starlight content in the first-stage fitting was 72.5\%, 
  and the last-stage fitting gave 65.6\% for a Lorentzian fit, and 62.8\% for a Gaussian fit (see Table~\ref{tab:nasze}). 
  We partially relied on the wide-range
  STARLIGHT results in selecting the dominant starlight components.
  We allowed them to vary and we performed an interactive fit of the two dominating components as well as of the total starlight normalization, 
  and all the other parameters, including both starlight and power-law extinction, and selected emission lines. Lines in the wavelength range 
  4400 - 5100 \AA \ are fitted, that is, masking this region was removed from the original masking scheme prepared for the STARLIGHT code. 
  The resulting wide-range fit is shown in the lower panel of Fig.~\ref{fig:broad_band}. The best Fe II template in this case was d11-m20-20-5, 
  broadened by convolution with a  Gaussian of 800 km s$^{-1}$ in width. The formal fit is better than that obtained from STARLIGHT. 
  The stellar content dramatically decreased, down to 3.4\%, that is, even more than in our narrow band fitting. 
  Thus, allowing for an extinction in the power law modifies the starlight strength very significantly.
  Extinction in the power-law component was $A_V = 0.24$, and extinction in the starlight was negligible. 
  Stellar content also changed from a dominating population of 25 Myr-old moderate-metallicity stars to equal contributions from 
  the 25 Myr-old population and a population of 3.1 Myr-old high-metallicity stars. In this fit the ratio $R_{Fe}$ is 0.94, slightly lower than given in Table~\ref{tab:summary}.
  We then limited ourselves again to the narrower band, but preserving the new starlight content and its relative normalization with 
  respect to the power law, and we also preserved the two values of extinction from the wide-range fit. 
  We then obtained a new value of $R_{Fe}$ equal to 1.27, higher than before. During the narrow-band fitting the normalization of the Fe II went up. 
  This is not surprising since the normalization of Fe II in UV and in the optical band are actually not correlated as seen in observations 
  (Kovacevic et al., 2015). Thus, the accuracy of the fitted values of $R_{Fe}$ is about 20\%, but the starlight component is still relatively uncertain.

\begin{figure}
 \centering
 \includegraphics[width=0.95\hsize]{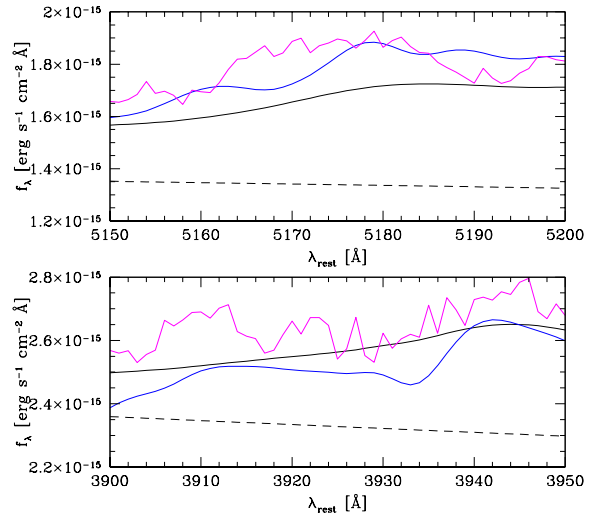}
 \caption{The Mg I and Ca II K absorption bands of the spectrum of SDSS130112.91+590206.6 (magenta line), the model based on automatic STARLIGHT run (blue line), and the model without the automatic use of the STARLIGHT code (black line). Dashed line shows the power-law position for the second model; in the first model the power law is much lower. Absorption features widths of a few \AA, corresponding to stellar dispersion and instrumental broadening, are expected at 3934 \AA~(Ca II K) and 5167, 5173, 5184 \AA~(Mg I triplet). They are not seen directly in the data, since in strong Fe II emitters absorption features are masked by strong Fe II contamination. Therefore, the starlight appears not to form the basis of better representation of stellar absorption, as was possible for REJ 1043+396 (Czerny et al. 2016).}
 \label{fig:absorption}
\end{figure}

The stellar content is most reliably measured when stellar absorption features are well visible and properly modeled.  For the source discussed
above, we thus determined which of the two solutions (two-step results and broad-band fitting) better reproduces the two most important stellar absorption features, Ca II K and Mg I. The fits in these two spectral regions are shown in Fig.~\ref{fig:absorption}. The data do not convincingly indicate a better solution. The signal-to-noise ratio (S/N) is not a problem, the source SDSS130112.91+590206.6 has a g magnitude  of 15.6, comparable to the source RE J1034+396 (g magnitude of 15.4) where the same absorption features were clearly visible (Czerny et al. 2016). The problem lies in the strong Fe II contamination. In the wavelength range 5100 - 5200 \AA~ there are 10 Fe II transitions in our templates, some of them strong, and 15 transitions are in the range 3900 - 4000 \AA. Using a single normalization for the whole Fe II template does not provide a precise fit to the data, and the possible absorption feature is buried in Fe II. Visual inspection of the Ca II K range seems to suggest that high starlight level is actually somewhat better. However, extension of the fits from 4400 - 5100 \AA~ to 4400 - 5200 \AA~ does not help. In the absence of well visible absorption features the fit relies in the broad band spectral shape. In their recent work on stellar kinematics in active galaxies Le et al. (2017) used dedicated MMT observations in the 5100 - 5400 \AA~ band supplementing their SDSS spectra.

\section{Conclusions}

We reanalyzed 27 spectra from the quasar catalog of Shen et al. (2011). The selected sources have the highest values of the Fe II to H$\beta$ and relatively low errors in these parameters. We improved the analysis by taking into account several Fe II templates, Balmer Continuum, and starlight contamination. Our analysis shows that:

(i) {Our values of the Fe II to H$\beta$ ratio are significantly lower than in the original Shen et al. (2011) catalog.}

 (ii) {Among the six most extreme objects, three objects have very narrow H$\beta$ lines and a FWHM between 1450 and 2100 km s$^{-1}$ . The other three objects have a FWHM(H$\beta$) above 4500 km s$^{-1}$ and therefore they do not seem to form a homogeneous sample, showing a broad range of Eddington ratios.}
  
(iii) {Sources with strong Fe II emission are very difficult to model reliably; particularly the determination of the stellar content and the [OIII] line intensity is complicated due to strong Fe II contamination.}

\section*{Appendix}
\subsection*{Comments on individual objects}

 Object SDSS125343.71+122721.5 has high $R_{Fe}$ in the Shen et al. catalog (see Table~\ref{tab:objects}) but negligible $R_{Fe}$ in our fits, and visual inspection also does not show traces of Fe II emission. We discussed this issue with the authors of the catalog, and apparently, this is a mistake in their automatic recording routine (Yue Shen, private communications). We therefore ignored this object when giving any mean/median values, but we plot the source for the record, and as an example of weak Fe II emission.

 Narrow [OIII] lines in object SDSS125100.44+660326.8 could not be fitted if assumed to be at the rest frame implied by the redshift taken from Shen et al. (2011). We considered the change of the source redshift but in this case the Fe II pseudo-continuum features, well visible in this source due to low broadening (only 700 km s$^{-1}$) were not reproduced at the proper positions. We thus allowed in this source for a redshift or a blueshift of both [OIII] lines,  and Fe II lines,  as we did in all our fits for the broad H$\beta$ component. We obtained much better fit for [OIII] lines shifted by 180 km s$^{-1}$ toward longer wavelengths. The best template was d11-m20-21-735 (i.e. particle density $10^{11}$ cm$^{-3}$, turbulent velocity 20 km s$^{-1}$, and ionization parameter, $\Phi$ $= 10^{21}$ cm$^{-2}$ s$^{-1}$).  Final reduced $\chi^2$ is now significantly lower (4.39). We thus obtained three values of the redshift in this source, depending on the line system used as a reference. Broad H$\beta$ line redshift is $z = 0.2823532$, [OIII] lines give $z = 0.2825588$, and Fe II implies the redshift $z = 0.28015$. If the [OIII] redshift is treated as a reference, then the Hbeta and Fe II emission are blueshifted by 62 km s$^{-1}$, and 723 km s$^{-1}$, respectively. The last value is large but still within the range given by Hu et al. (2008), and consistent with the trend between the shift and the Eddington ratio. 

   These redshifts can be compared with the redshifts available for this source in the literature. The value $z = 0.282$ comes from Schneider et al. (2010), and it was determined from multi-line fits, with a typical error of 0.004, and a possible systematic error of 0.003 (Hewett \& Wild 2010). The same redshift was used by Shen et al. (2011), and at the stage of starlight fitting by us. Wild \& Hewett (2010) in their DR7 Value Added
   catalog\footnote{http://classic.sdss.org/dr7/products/value\_added/index.html\#quasars}
   give $z = 0.282663$, at the basis of [OIII] line fitting, and the same value is later used in DR12 catalog\footnote{\scriptsize{http://skyserver.sdss.org/dr12/en/tools/explore/Summary.aspx?sid=557377386966968320}}. Our redshift based on the same line is close to this value; the difference is equivalent to 31 km s$^{-1}$.

Fits also improved when we allowed for [OIII] line shifts for two other objects: SDSS144645.93+403505.7 and SDSS152939.29+203906.8. Particularly in this last case the required [OIII] shift, towards shorter wavelengths this time, was as high as 480 km s$^{-1}$, but $\chi^2$ dropped from 2.75 down to 1.25. The source also required large broadening of the Fe II template, 1600 km s$^{-1}$, and the favored template was d11-m20-20.5-735.dat (i.e. particle density $10^{11}$ cm$^{-3}$, turbulent velocity 20 km s$^{-1}$, and ionization parameter, $\Phi$ $= 10^{20.2}$ cm$^{-2}$ s$^{-1}$).   

 Object SDSS165252.67+265001.9 also required significant Fe II broadening (1400 km s$^{-1}$), and the highest value of the turbulent velocity (30 km s$^{-1}$, template dd11-m30-20.5-735. In general, sources with broad H$\beta$ line required stronger broadening of the Fe II template. This connection was already noted by Hu et al. (2008).

 Object SDSS133005.71+254243.7 has moderate $R_{Fe}$ but an extremely broad H$\beta$ line. We thus refitted this object more carefully, searching for the best Fe II template among Bruhweiler \& Verner (2008) options and best broadening. In this object the Fe II pseudo-continuuum also required extreme broadening of 2300 km s$^{-1}$, and the best fit template was d11-m10-20-5 (i.e., particle density $10^{11}$ cm$^{-3}$, turbulent velocity 10 km s$^{-1}$, lower than typical, and ionization parameter, $\Phi$ $= 10^{20.5}$ cm$^{-2}$ s$^{-1}$). This template also gives a strong and broad contribution close to the [OIII] line which is clearly visible in the spectrum. Achieved $\chi^2$ is $870/700 \ d.o.f$.  The $R_{Fe}$ ratio is 0.9, somewhat higher than obtained from systematic fitting of all objects but nevertheless smaller than $1.93$ reported by Shen et al. (2011). Still, the Fe II is not yet well modeled in the region $\sim 4550$ \AA \ (see Fig.~\ref{fig:broad_line}). Still better fits can be achieved modeling individual Fe II transition groups as done by Kovacevic et al. (2010). However, the accuracy of the EW(FeII) computation is not strongly affected. We compared the computed EW(FeII) from the fitted Fe II model and from the data by subtracting all other components, and the difference was by 1.4 \%. We also tried a different approach: we used only the strongest transitions from the d11-m10-20-5 template (above $10^6$ in units of Bruhweiler \& Verner 2008), there were 23 such transitions in the fitted region, and we modeled Fe II as a sum of 23 Gaussians, with the relative normalizations as in the template. Then we allowed some of these normalizations to vary (two transitions close to 4500 \AA, and a transition at 5031.87 \AA), and we reached a satisfactory $\chi^2$ of 739, with H$\beta$ still modeled as a single Gaussian. The resulting $R_{Fe}$ from this fit was 1.05. Thus, our modeling, independently from the method, is roughly within 20\% accuracy.

The starlight content in this object is high,  40.5\%. This, combined with very broad line might suggest a relatively high inclination of the source. Lack of asymmetry in H$\beta$ suggests that the source does not show any extreme phenomena, such as massive outflows seen in some quasars (e.g., Maddox et al. 2017). Power law extinction is moderate in this source, $A_V = 0.297$, which causes the decrease of the flux by a factor 0.695 at 4400 \AA \ and 0.741 at 5100 \AA, so the curvature in the power law component is not considerable in this restricted wavelength range. In the case of this source we also tried the extinction curve of Czerny et al. (2004). However, in the limited range of 4400 \AA~ --- 5100 \AA~ where we use this law, the change is small (decrease of flux by a factor of 0.76 at 4400 \AA~ and 0.81 at 5100 \AA), so fits are unaffected. 


\begin{acknowledgements}
 We are deeply grateful to the anonymous referee for multiple comments on the manuscript. The project was partially supported by Polish grant No. 2015/17/B/ST9/03436/.
The STARLIGHT project is supported by the Brazilian agencies CNPq, CAPES and
FAPESP and by the France-Brazil CAPES/Cofecub program. The Fe II theoretical templates described in 
Bruhweiler \& Verner (2008) were downloaded from the
web page http://iacs.cua.edu/personnel/personal-verner-feii.cfm with the permission of the authors. 
This research has made use of the NASA/IPAC Extragalactic Database (NED) which is operated by the 
Jet Propulsion Laboratory, California Institute of Technology, under contract with the National Aeronautics and Space Administration.
\end{acknowledgements}

\end{document}